# Understanding the dynamics emerging from infodemics: A call to action for interdisciplinary research

Stephan Leitner
University of Klagenfurt
Department of Management Control and Strategic Management
stephan.leitner@aau.at

Bartosz Gula
University of Klagenfurt
Cognitive Psychology Unit
bartosz.gula@aau.at

Dietmar Jannach
University of Klagenfurt
Department of Applied Informatics
dietmar.jannach@aau.at

Ulrike Krieg-Holz
University of Klagenfurt
Department of German Studies
ulrike.krieg-holz@aau.at

Friederike Wall
University of Klagenfurt
Department of Management Control and Strategic Management
friederike.wall@aau.at

Abstract

Research on infodemics, i.e., the rapid spread of (mis)information related to a hazardous event, such as the COVID-19 pandemic, requires the integration of a multiplicity of scientific disciplines. The dynamics emerging from infodemics have the potential to generate complex behavioral patterns. In order to react appropriately, it is of ultimate importance for the fields of Business and Economics to understand the dynamics emerging from it. In the short run, dynamics might lead to an adaptation in household spending or to a shift in buying behavior towards online providers. In the long run, changes in investments, consumer behavior, and markets are to be expected. We argue that the dynamics emerge from complex interactions among multiple factors, such as information and misinformation accessible for individuals and the formation and revision of beliefs. (Mis)information accessible to individuals is, amongst others, affected by algorithms specifically designed to provide personalized information, while automated fact-checking algorithms can help reduce the amount of circulating misinformation. The formation and revision of individual (and probably false) beliefs and individual fact-checking and interpretation of information are heavily affected by linguistic patterns inherent to information during pandemics and infodemics and further factors, such as affect, intuition and motives. We argue that, in order to get a deep(er) understanding of the dynamics emerging from infodemics, the fields of Business and Economics should integrate the perspectives of Computer Science and Information Systems, (Computational) Linguistics, and Cognitive Science into the wider context of economic systems (e.g., organizations, markets or industries) and propose a way to do so. As research on infodemics is a strongly interdisciplinary field and the integration of the above-mentioned disciplines is a first step towards a holistic approach, we conclude with a call to action which should encourage researchers to collaborate across scientific disciplines and unfold collective creativity which will push forward research on infodemics substantially.

## 1. Introduction

In the past few months, we have all witnessed the short-term responses to the COVID-19 pandemic, such as extensive lockdowns in both the private and the professional spheres. We have experienced a multiplicity of near-term consequences of such a hazardous event, for example, the enormous costs in lives and the substantial increase in uncertainty related to the capacity of healthcare systems. Alongside the COVID-19 pandemic, we can also observe an accompanying *infodemic*. Infodemics are referred to as the rapid spread of information and misinformation that accompanies a pandemic, which makes it difficult for decision-makers to find reliable sources of information (Vaezi and Javanmard 2020; Zarocostas 2020). *Infodemiology* (i.e., the research area related to infodemics) is concerned with the determinants and the distribution of information and misinformation among decision makers, with the transition of information in terms of potential knowledge gaps between best evidence (e.g., expert knowledge) and what most people do or believe, and markers for what can be regarded high-quality information (Eysenbach 2002; Zannettou et al. 2019). Recently, it can be observed that the field of Infodemiology not only focuses on the supply side of information, in terms of what sort of information is available, but also takes into account the demand side, in terms of what kind of information is searched for by decision makers (Eysenbach 2020).

In the context of the COVID-19 pandemic, we can observe that infodemics often have negative effects on individual behavior and on the efficiency of counter-measures deployed by policy-makers (World Health Organization 2018): A prominent example for the rapid spread of (mis)information is the extensive promotion of hydroxychloroquine as treatment for the SARS-CoV-2 virus in social media without reliable evidence of its efficacy and secondary effects, which was followed by an emergency use authorization by the US Food and Drug Administration (FDA) and a significant increase in prescriptions (Baker et al. 2020a; Lenzer 2020). Later, this authorization was revoked by the FDA as no benefit for decreasing the likelihood of death or recovery by treating patients with hydroxychloroquine could be observed, whereas serious side effects such as heart rhythm problems, kidney injuries, and liver problems were reported (U.S. Food & Drug 2020). Interestingly, the extensive promotion of hydroxychloroquine was state-sponsored misinformation which has a tendency to propagate faster and further than other types of misinformation (Zannettou et al. 2019). In the context of the COVID-19 pandemic, we can, however, observe that non-state-sponsored information also leads to adaptations in individual behavior (Kim et al. 2019): Rumors about a possible lockdown in the region of the Lombardy (before the lockdown was officially announced by the government), for example, led to overcrowded trains and airports in order to escape this region (Cinelli et al. 2020). Illyas (2020) and Ahmad and Murad (2020) report that the misinformation disseminated in social media has increased panic and fear significantly, which resulted in a shift in household spending. Their finding is in line with Hall et al. (2020), who argue that spending in some sectors has increased as a response to COVID-19 (e.g., stockpiling of needed home goods), while other sectors experience a sharp decline in spending, such as the hospitality sector and the retail industry, which might also be caused by a tendency to postpone purchases and consumption of discretionary products (see also Baker et al. 2020c; Sheth 2020). Addo et al. (2020) and Chaudhary (2020) further argue that the increased panic and fear have also led to a paradigm shift of buying behavior towards online shopping. These short-term changes in individual behavior are likely to have a tight connection with issues of personal protection, such as social-distancing, the obligation to wear face-masks in (selected) public areas, and further protective measures (Funk et al. 2009).

Aside from these short-term consequences, long-term effects can be expected in the future (Donthu and Gustafsson 2020). These long-term effects include major impacts on organizations and markets which might, for example, be caused by adaptations in consumer and travel behavior (Fernandes 2020; Nicola et al. 2020; Leitner 2020). As the pandemic and the infodemic are still ongoing, the scale of the effects, however, is difficult to estimate (Baker et al. 2020b; Goodell 2020). Gillingham et al. (2020) and Donthu and Gustafsson (2020) analyze potential long-term effects of the COVID-19 pandemic for the fields of Business and Economics and expect such effects in (i) investments, (ii) consumer behavior, and (iii) markets. Related to (i) investments, Jordà et al. (2020) take a historical perspective and conclude that in the aftermath of past pandemics, returns on assets have dropped and reduced economic growth could be observed as a consequence of reduced investments. Contrary to Gillingham et al. (2020) they argue that it is not at all certain that we will observe similar long-term effects of the COVID-19 pandemic since the underlying conditions are different. Avoiding investments and saving capital instead, for example, means (almost) negative returns nowadays. With respect to (ii) consumer behavior, Sheth (2020) argues that some habits will fall back to what was regarded to be normal before the recent pandemic, but also adds that it is inevitable that some habits related to consumption will change. Gillingham et al. (2020) provide an illustrative example of changing habits: They argue that – also in the long run – individuals might switch to modes of private transport, if they remain fearful of taking public transportation which, as a consequence, might result in new framework conditions for the entire transportation sector. Regarding the effect of the pandemic for (iii) markets, Donthu and Gustafsson (2020) presume that many industries will be affected in the long run, such as the travel industry which already struggles with empty hotel rooms and airlines which already cut workforce by significant numbers. They also argue that organizations will rethink their supply chains and move them closer where they are needed, which is likely to have serious consequences for the fields of Business and Economics as well.[1]

Previous studies suggest that some shifts in both short- and long-term behavior, as outlined above, are driven by false beliefs (Bernstein et al. 2005; Geraerts et al. 2008), i.e., by an individual belief or representation about the world which does not match reality (Bauminger-Zviely 2013). Infodemics play an important role in this context as they substantially contribute to the formation and propagation of false beliefs: It is well known that repeating information increases its subjective truth (Hasher et al. 1977; Unkelbach and Rom 2017). The frequent sharing of misinformation about COVID-19 and the multiplication of this information via social media are, for example, likely to produce false beliefs (see also Unkelbach et al. 2019). Previous studies also show that language plays an important role in the development of false beliefs: For instance, Farrar and Maag (2002) elaborate on the importance of receptive vocabulary in the context of false belief formation, and Slade and Ruffman (2005) find that over-all language comprehension rather than any specific aspect (e.g., linguistic complexity) impacts on false beliefs. With respect to complementation and false beliefs, De Villiers and Pyers (2002) argue that sentential complement structures have a unique role in false-belief understanding; the mental state verb "think", for example, allows to embed one proposition into another and, as a consequence, to distinguish between what a person thinks and what is true (see also Farrar et al. 2013). Loftus (2005) highlights the severe effect of infodemics when beliefs are formed: She shows that misinformation can even lead to the creation of recollection of past events that have *never* occurred. What makes false beliefs particularly challenging is that they are not only likely to have immediate behavioral (short-term) consequences, but may also trigger long-term behavioral consequences, as they tend to last (Laney et al. 2008). Once formed, false beliefs might also be difficult to convert into true beliefs: Niiniluoto (1977, 2011) shows that even true information can lead to faulty belief revision, so that the revised belief is even farther from the truth than the unrevised one.

---

[1] For a detailed discussion of potential effects of the ongoing pandemic for different areas of Business and Economics, the reader is also referred to the thematic collection on "COVID-19 impact on business and research" published in volume 117 of the Journal of Business and Research.

In order to react appropriately to the current situation, it is of ultimate importance for the fields of Business and Economics to understand the dynamics which emerge from infodemics: First, from a short-term perspective, the interactions between policy measures taken in order to contain the pandemic, the rapid spread of information and misinformation, the formation of beliefs and their immediate behavioral implications are widely unknown. Second, from a long-term perspective, research on the interaction between infodemics, belief formation and revision and long-term behavioral implications is required. Third, as the pandemic is still ongoing, the interaction of behavioral implications on different timescales cannot be ignored: It is, for example, widely unknown, how interactions between concurrently emerging short- and long-term consequences caused (or enforced) by infodemics are shaped. Fourth, from a macroscopic point of view, it is of ultimate interest to understand how all these consequences of infodemics accumulate to behavioral patterns of organizations, markets, industries, economies, or societies. The ongoing digitalization adds additional complexity to infodemics-related research, as the popular use of digital communication technologies and social media accelerates the diffusion of information substantially: This trend is also reflected in the spread of (mis)information related to the COVID-19 pandemic (Pulido et al. 2020; Rovetta and Bhagavathula 2020); it has led to significantly more worldwide fear, panic, and uncertainty when compared to infodemics that emerged from previous pandemics (Cinelli et al. 2020; Vaezi and Javanmard 2020). These observations highlight the need for the fields of Business and Economics to develop a deep(er) understanding of the dynamics related to infodemics. We should make an effort to learn from the current pandemic and infodemic in order to be well-prepared for future outbreaks (Donthu and Gustafsson 2020).

Dealing with infodemics requires a highly interdisciplinary approach and this position paper is a call to action for such kind of research on infodemics. As already mentioned, according to Eysenbach (2002, 2020), the field of Infodemiology is concerned with both, the supply- and the demand-side of information. In order to give credit to the importance of interdisciplinary collaboration on this topic, we particularly take into account four disciplinary perspectives which we consider particularly important for representing the perspectives on information supply and demand, and which, not at least through their interplay, are very promising to substantially contribute to gaining a deep(er) understanding of the dynamics emerging from infodemics. First, through the lens of Computer Science and Information Systems, the information accessible to individuals is central, whereby the way information spreads in a society is strongly affected by the employed algorithms for information provision and by the way information provision is aligned with personal interests. The applied algorithms are often designed to react to the demand of specific pieces of information, which is why Computer Science and Information Systems is not only concerned with information supply but also connects the perspectives of supply to demand (see Sec. 2.1). Second, from the perspective of (Computational) Linguistics, specific language signals in social communication which emerge during pandemics need in-depth investigation (e.g., emotion-related words, avoidance of causal terms). Incorporating linguistic patterns in the context of infodemics is highly relevant since these patterns strongly affect the way information is packaged, interpreted, fact-checked, made sense of by non-expert persons, and how misinformation is automatically detected. The field of (Computational) Linguistics puts strong emphasis on the supply side of information and contributes to accurate translation of knowledge to socially spread (mis)information, thus allowing (automated) fact-checking and the identification of linguistic patterns which might distort the message contained in the information (see Sec. 2.2). Third, from a Cognitive Psychology point of view, the focus is on the assessment of how available information (e.g., driven by algorithms and by personalization) and the language signals affect the way pandemics and policies are perceived (e.g., in the form of false beliefs) and on how they influence individual behavior. This field, therefore, is mainly concerned with the demand side of information (see Sec. 2.3). Fourth, while the Cognitive Psychology viewpoint mainly focuses on the level of the individual, the perspective of Business and Economics allows for integrating the aforementioned perspectives into the wider context of economic systems by analyzing how individual behavioral implications (which probably materialize on multiple time-scales) interact with rapidly spreading (mis)information and with each other, and how they accumulate to macroscopic patterns of organizations, markets, industries, economies or societies (see Sec. 2.4). By doing so, the holistic and integrated perspective resulting from the collaboration of the fields mentioned above, prepares the ground for gaining fundamental insights into the dynamics emerging from infodemics.

## 2. Selected disciplinary perspectives on infodemics and the potential of their integration

### 2.1. Computer Science and Information Systems: Algorithm-controlled information dissemination and personalization

In the last two decades, we have observed disruptive changes in terms of how information is spread in societies. While traditional mass media channels continue to exist, the Web and in particular Social Media have become the main source of information for the majority of people in today's digital society. These developments have made information not only more easily accessible for consumers but have also led to the rise of digital businesses like Google, Facebook, or Twitter, which have an enormous, global-scale reach. As a result, information that is spread via such channels is often viewed by millions of people within short periods of time.

Unlike traditional media, such online information channels do no longer rely on classical gatekeepers, such as journalists or editors, to ensure the quality or reliability of the information that is spread. This makes such channels generally vulnerable of being used for the dissemination of misinformation in the form of an infodemic. A particular problem that can additionally contribute to the rapid spread of misinformation on such channels is grounded in the fact that the selection of the content provided through these channels is determined by algorithms. These algorithms are usually optimized to identify, in a personalized way, those pieces of content that are most likely viewed by an individual consumer. The underlying reason for trying to optimize the number of 'clicks' instead of finding the most relevant pieces of content for a user lies in the advertisement-based business model of many online services.

The algorithms that select and rank the contents for the individual users, e.g., in the form of recommender systems (Jannach et al. 2010), are in most cases based on statistics and machine learning. As such, they learn over time which pieces of content, e.g., news articles or videos, are likely to be clicked on by users. As a result of using such learning strategies, a reinforcement ("blockbuster") effect can often be observed (Fleder and Hosanagar 2009) where the rich get richer: Content that has reached a certain level of popularity continues to be recommended to even more users, which, in turn, can propagate the formation of false beliefs.

But this reinforcement bias is not the only problem of today's algorithm-controlled dissemination of information. The increased levels of personalization can also lead to filter bubbles (Pariser 2011) and echo chambers (Celis et al. 2019). The reason is that modern algorithms learn over time which type of content a certain user likes or dislikes. Again, to optimize their business, the algorithms will try to focus on content that the user is likely to click on. In the context of an infodemic, this can lead to the effect that users at some stage are presented only with one of several existing theories, e.g., regarding the appropriateness of certain measures taken by policymakers to contain the pandemic. This one-sided information state, as a result, can further reinforce the spread of misinformation and strengthen false beliefs.

In the academic literature in the fields of Computer Science and Information Systems, these potentially negative effects of personalization and algorithm-driven content recommendations have recently attracted increased research interest. Correspondingly, questions of algorithmic "fairness" and how to provide users with understandable explanations why certain items were recommended moved into the focus of researchers. These problems are, however, far from being solved. Besides algorithmic challenges, a fundamental problem usually lies in defining what being fair actually means in a given context (Friedler et al. 2016; Burke 2017; Abdollahpouri et al. 2020).

### 2.2. Linguistics: Signals and patterns of (mis)information in social communication

While Computer Science algorithms distribute containers of information fitting systematic behavioral patterns (e.g., counting clicks as indicators of content-preferences and recommending "similar" containers), the study of language (Linguistics) is directly concerned with the contents of these information containers – at the lexical level of words, at the compositional level of utterances (assertions, claims, stipulations, etc.) and at the pragmatic level of social language use (e.g., information exchange, stance taking or persuasion in debates, deceptive, aggressive and other kinds of "toxic" speech). Hence, from a linguistic viewpoint, the analysis of infodemic utterances (including fake news, propaganda, troll messages, etc.) focuses primarily on linguistic signals indicating their factuality, claim, opinion, trust, or believability status. Early work on distinguishing liars from truth-tellers tried to identify such cues directly, e.g., liars using fewer self-references and more other-references, avoiding causal terms, using more negative emotion words, and revealing characteristic syntactic patterns (Burgoon et al. 2003; Newman et al. 2003; Hancock et al. 2007; Lee et al. 2009). Yet, a recent survey by Gröndahl & Asokan (2019) provides ample evidence that the search for *general* stylistic traces of deception, i.e., linguistic markers carrying high emotional load, a high degree of generality/abstractness, high/low use of first-person pronouns, high use of verbs and certainty-related words, might miss significant portions of such biased language.

An alternative avenue of research on testing the believability of assertions does not focus on direct linguistic cues but has its roots in ancient rhetorics (e.g., considering syllogisms) and concentrates on discourse pragmatics. This approach is based on the seminal work of Stephen E. Toulmin (1958). He introduced the so-called Toulmin schema which distinguishes three fundamental and three auxiliary components of coherent argumentative discourse. According to Toulmin, a claim (thesis, conclusion) can be derived from a piece of information (data) by making use of inference rules (warrants). This standard deductive reasoning scheme, well-known from standard formal logics, can be further complemented by *supporting evidence* (backing, typically general norms, value sets, moral standards, etc.) adding additional evidence mostly to warrants but also to data. The strength or certainty of information can further be adjusted by modal qualifiers at any stage (such as "mostly", "probably"), whereas exclusive conditions (rebuttals) can be expressed to indicate exceptions to general rules. Due to its high degree of idealization, the Toulmin schema has stimulated research in many fields – from formal logics and artificial intelligence (Verheij 2009; Caminada 2018) to pragmatics-focused linguistics (van Eemeren and Grootendorst 2004) and computational linguistics (Cabrio and Villata 2012; Hidey et al. 2017). In particular, the automatic recognition of (im)proper argumentation structure has recently received enormous attention as a side effect from argumentation mining (Lippi and Torroni 2016; Habernal and Gurevych 2017). There are also strong links of argumentation process modeling to the fields of software agents (Parsons et al. 1998) and multi-agent decision making (Karacapilidis and Papadias 1998).

Rather than explicitly enumerating specific linguistic cues, current work in the field of automatic natural language processing investigates the potential of automatic classification methods to distinguish between truth-friendly and -unfriendly language use (for a recent survey, cf. Fitzpatrick et al. 2015), employing deep learning architectures, in particular (Popat et al. 2018; Liu and Wu 2020). Especially, the areas of fake news detection (Popat et al. 2018; Liu and Wu 2020) and fact/claim checking (Rashkin et al. 2017; Volkova et al. 2017; Thorne and Vlachos 2018) have recently generated substantial insights into tracking the truth status of verbal statements.

However, unlike the vast majority of language understanding tasks, (non-expert) persons face substantial problems in discerning deceptive from non-deceptive language. Actually, their accuracy – when it comes to detecting textual deception – is approximately on a chance level, or even worse (Bond and DePaulo 2006) so that valid ground truth is hard to attain. Hence, besides testing the performance of deception/fake classifiers the creation of reliable test data sets (gold standards) is a major challenge in current language-focused infodemics research and covers a broad range of topics: deception and lies (Fitzpatrick and Bachenko 2019), fake news, and fact checking of claims (Augenstein et al. 2019). They form the benchmarks underlying specialized challenge competitions, e.g., aiming at fact checking in social media (Barrón-Cedeño et al. 2020).

**2.3. Cognitive Psychology: Risk perception, false beliefs, and individual behavior**

The Cognitive Psychology perspective switches the focus from the way information is designed and disseminated to how pandemics and infodemics are perceived and how beliefs are formed and revised. Research on decision-making and thinking distinguishes between two modes of thinking: One more deliberate and analytic, the other more intuitive and heuristic (Kahneman 2003; Evans and Stanovich 2013). Intuitive thinking, though often accurate, may lead to systematic biases that appear even more likely in the case of information of low quality or systematic misinformation spreading through social media. For example, people prefer to search for information that supports rather than challenges prior beliefs (confirmation bias) and are overly reliant on the believability of conclusions when assessing the quality of arguments (belief bias). Also, the extent to which one's own opinions are shared by others is frequently overestimated (false consensus effect), as is the impression of how widespread beliefs of others are in specific social network structures (majority illusion; Lerman et al. 2016). Various cognitive tools have been tested in recent years that aim to increase individuals' ability to judge the quality of information and the credibility of sources, and to empower autonomous decision-making in general (Lorenz-Spreen et al. 2020).

Despite these organized efforts of misinformation containment and public fact checking, false beliefs tend to persist. Belief updating and revision occurs rather slowly, following the principle of minimal change that does not question the wider belief system or more fundamental core beliefs (Gärdenfors 1992). Lewandowsky et al. (2012) give an overview of the causal mechanisms which help explain why people often do not change their beliefs and behavior even if the information on which their beliefs are based has been shown to be false and retracted. When seeking information, people appear to accept the truthfulness of speakers by default and tend to question the quality of information mostly if it is inconsistent with their beliefs, produces an incoherent narrative, or overt cues or significant others question the credibility of the source. Information is usually part of broader narratives used by everyone to extract meaning from unfolding events. Mere fact checking is more likely to be effective, if it also offers verified alternative narratives for both, the original false and the verified information (Lewandowsky et al. 2012).

In a digital environment that develops at a fast pace, risk and digital information literacy are more important than ever. Research shows that risk perception may be distorted if explanations are missing, when actually accurate information is presented in a format that is not well-tuned to the cognitive system of recipients, or when the risks involve low-probability events with particularly threatening consequences (Slovic 2010; Gigerenzer 2015). For instance, people have difficulties in understanding what it actually means when today's weather report announces a 30% chance of rain, or when a public health organization posts that the relative risk of thrombosis increases by 100% when taking a contraceptive pill. Serious incidents such as the terrorist attacks on September 11 and perhaps COVID-19 incite strong emotions and increase perceived risks way beyond their actual probability. Gigerenzer (2004) showed that in the three months following September 11, American citizens reduced air travel which in turn led to a substantial increase in fatal car accidents. In order to reduce such hazards, public policy making seems to face the difficult task to not only consider direct, objective risks but also to take factors into account that influence public risk perception and to predict the corresponding behavioral consequences.

In order to increase risk and digital information literacy and autonomous decision-making in general, a number of cognitive tools have been introduced in recent years. Kozyreva, Lewandowsky, and Hertwig (2019) distinguish nudging and boosting. Whereas nudges typically target the design of the choice environment and require little activity from users, boosts address lasting changes in competences of users and require active cooperation. As an example for boosts, the procedure used by professional fact-checkers was translated into training interventions and decision aids that guide users in how to evaluate the credibility of the source, the evidence in arguments and how to read laterally, i.e., cross-check information on other sites (Wineburg and McGrew 2019). Another type of boost has been termed knowledge-based, deliberate ignorance (Hertwig and Engel 2016). With an overwhelming amount of easily available information, we routinely decide which information to pick up and which to ignore. Tools supporting deliberate ignorance make use of ratings or other cues to source quality. People seek information to satisfy basic needs such as the need for competence, the need for autonomy and the need to belong (Deci and Ryan 2000). Both public policies to cope with the infodemic and cognitive tools designed to improve decision-making autonomy, appear most promising the more they consider all three needs together.

### 2.4. Business and Economics: Integrating micro- and macro-perspectives on infodemics and coordination of individual behavior

From the perspective of Business and Economics, the three views discussed before can be integrated into the wider context of entire economic systems, such as organizations, markets, industries, economies or societies. From a macroscopic perspective, such systems can be regarded as complex and adaptive social systems from which some peculiar properties can be abstracted (Thurner et al. 2018): First, at the micro-level, such systems consist many entities with often complex interactions between them. These entities can, for example, represent consumers in a market, individuals who communicate information in their social networks, firms within an economy, entities which provide, personalize and disseminate information, institutions which provide automated fact-checking of information, policymakers, and the government. The interactions between the parts of a complex and adaptive system are driven by specific laws and channels of interaction. Moreover, interactions may be subject to adjustments and emergence shaped by the entities' actions and interactions. Against the background of infodemics, fundamental channels of interactions are, for example, social media platforms, further means of digital communication, personal interaction and other entities which provide individuals with personalized information.

Second, all entities of the micro-level of such a system have individual characteristics which cover, for example, roles to play in the society, states of information, linguistic patterns in communication, cognitive capacities to form and revise beliefs, the capabilities to fact-check information, and algorithms to automatically detect misinformation. Together with the laws and channels of interactions, the individual characteristics infer the macro-properties of the entire system. In the context of the COVID-19 pandemic, resilience to misinformation and information literacy, or specific types of herd behavior in terms of opinion (or belief) dynamics could be among such properties at the macroscopic level. From the fields of Business and Economics, short- and long-term behavioral implications of infodemics, such as adaptations in travel or buying behavior, changes in consumption patterns or adaptation of investment behavior could be of ultimate interest.

Third, the society adapts to its environment over time, whereby the adaptive change is substantially governed by the system's environment and the information accessible to the parts of the system (Simon 1990; Thurner et al. 2018). In the context of hazardous situations such as the COVID-19 pandemic, the system's environment is - in large parts - structured by policy-decisions: Enforceable (behavioral) rules such as compulsory face masks, travel bans, and lockdowns of entire industries shape the limits of a society's room for adaptivity substantially and might govern individual behavior into a specific direction. Previous research has indicated that these features of the environment tend to shape the short-term behavioral implications (Funk et al. 2009). In addition to the structure of the environment, such a system's behavior is affected by the states of information of its members, which can be interpreted as a function of the infodemics, their capabilities to collect, decipher, fact-check and make use of this information, and the resulting (false) beliefs which, in turn, are regarded to be a main consequence for long-term behavioral implications.

Integrating the perspectives of Computer Science and Information Systems, (Computational) Linguistics, Cognitive Psychology, and Business and Economics using a complex and adaptive system approach, as proposed in this paper, appears to be particularly fruitful:

- First, characteristics of (human) entities at the micro-level are in large parts shaped by factors which are fed by insights from these fields; such individual characteristics include cognitive capacities to form beliefs, fact-checking skills, and mandatory rules but also a wide range of communication strategies to properly situate linguistic behavior in a given discourse context. Further, non-human entities at the micro-level can, for example, represent algorithms to automatically fact-check information or recommend information to human entities. There might be complex dynamics at the micro-level: Non-human entities, on the one hand, usually acquire and adapt their capabilities over time, i.e., automated fact-checking might be carried out by intelligent software which autonomously learns to classify information, and information filtering algorithms collect personalized information, whereby characteristics of information recipients (such as personal interests) are to be learned over time. Humans, on the other hand, acquire and adapt their skills to efficiently fact-check information over time, they develop individual and situation-specific verbal communication routines, and form and revise (false) beliefs which induce long-term behavioral implications. Individuals might also self-organize themselves according to accessible information and their beliefs: Previous research, for example, shows that a collective identity (e.g., in the form of shared beliefs) facilitates the formation of groups (Van Dyke and Amos 2017), which might affect the behavioral implications induced by infodemics. At the micro-level, the consequences of interactions among these factors are still widely unknown. Integrating them in the framework of a complex and adaptive system promises to provide substantial insights into the behavioral dynamics of infodemics.
- Second, since pandemics and infodemics are continuous processes, effects on different timescales might be effective concurrently. For example, long-term behavioral implications such as the change in consumption habits or a shift in investment behavior might endogenously emerge, while new short-term behavioral adaptation due to issues of personal protection materializes. It is currently widely unknown how responses to pandemics and infodemics on different timescales interact and how the resulting behavioral implications for the fields of Business and Economics are shaped.
- Third, we can observe that the perception of the current pandemic and (individual) responses to it in both the long and the short run tend to be rather diverse (Dryhurst et al. 2020; Shefrin 2020), with diverse, heterogeneous and possibly dynamic behavioral implications at the level of the individual. The integration of the disciplinary perspectives considered in this position paper into a holistic approach can help us understand which macroscopic patterns emerge from the microscopic dynamics exemplified above. Understanding both the micro- and the macro patterns will provide the fields of Business and Economics substantial insights and will aid in efficiently and effectively responding not only to the current but also to future infodemics.

One particularly interesting question in this context is, how individuals can be guided in their adaptive behavior so that both the micro- and the macro-properties resemble required characteristics, such as information literacy or specific types of behavior to mitigate negative effects of the pandemic for organizations or the economy.[2] The behavioral dynamics emerging from infodemics might add additional complexity to this question as coordination requires communication, and the efficiency of the communication might be affected by infodemics.

Infodemics and coordination of individual and collective behavior towards a specific set of objectives are interrelated in a multiplicity of ways. First, one may argue that infodemics hinder coordination of behavior towards a specific objective: misinformation might unfold in unwanted behavioral dynamics as it might provide individuals with reasons for not adhering to specific policy decisions or for behaving so that specific objectives cannot be achieved. Believers in conspiracy theories, for example, may regard the actions taken by authorities as evidence for their theory and, thus, refuse even more to comply with policies. Through the lens of Business and Economics, such adverse behavior might be of ultimate interest: Understanding the behavioral dynamics which emerge from infodemics are, for example, of high importance in a planning context; understanding the dynamics allows for better anticipating the behavior of an organization's stakeholders, which might help avoiding a substantial amplification of uncertainty in the aftermath of COVID-19.

Second, certain efforts towards coordination may be made to mitigate infodemics. In the short term, such counter measures could be the design of algorithms for information provision, personalization of provided information, public fact-checking of information or the automated detection of misinformation by the means of computational linguistics. In the long term, coordination efforts could, for example, cover issues of education in order to increase risk and digital information literacy (e.g., in terms of boosting). In addition to gaining a mere understanding of the dynamics emerging from an infodemic (as discussed above), efficient coordination enables guiding the behavior. In order to do so, it is utterly important to understand the dynamics emerging from infodemics in terms of how individuals and collectives respond to coordination efforts. Policies to strengthen the economy in the aftermath of COVID-19, for example, need to be carefully tested before being launched in order to avoid adverse effects triggered by unprecedented dynamics emerging from the COVID-19 pandemic and the accompanying infodemic.

Third, as coordination involves communication, the question whether communication in the course of coordination could accidentally or deliberately also be part of the infodemic is particularly difficult. Therefore, it is of ultimate importance to understand the dynamics which an infodemic unfolds so that the communication in the course of coordination does not fuel unwanted behavioral, social and economic dynamics but contributes to the guided self-organization of a society so that intended patterns (e.g., in terms of achieved objectives) emerge (Wall 2019; Leitner and Wall 2020; Reinwald et al. 2020).

---

[2] In a broad sense of meaning, coordination can be referred to as the manifold of mechanisms to guide self-organization which aim at aligning the behavior of individuals to specific objectives (Baldassari 2009).

## 3. Summary and outlook

In this position paper, we particularly focus on the phenomenon of infodemics, i.e., on information and misinformation which spreads rapidly in developed societies and makes it difficult for decision-makers to find reliable, trustworthy sources of information. The dynamics emerging from infodemics can be complex, result in adverse individual behavior and might render policies to contain the effects of hazardous events inefficient. Such an infodemic can be observed in the context of the ongoing COVID-19 pandemic. We argue that it is of ultimate importance for the field of Business and Economics to gain a deep(er) and substantial understanding of the dynamics emerging from the current infodemic. We propose to employ powerful techniques like agent-based modelling which could be a valuable approach for predicting the effects of actions taken in a crisis situation such as the COVID-19 pandemic in conjunction with the human behavior within the crisis (Adam 2020; Squazzoni et al. 2020)

Research on infodemics is a strongly interdisciplinary endeavor: It requires the integration of a multiplicity of disciplines. In this position paper, we put a particular focus on the fields of Computer Science and Information Systems, (Computational) Linguistics, Cognitive Psychology, and Business and Economics. The field of Business and Economics has the potential to integrate the views of these disciplines into a holistic research perspective. We argue that this can be done by taking a complex and self-adaptive system approach: Other disciplines often focus on the level of the individual, for example in terms of individual characteristics such as cognitive capacities, information search behavior, or mandatory/optional use of verbal communication rules and strategies. The integration into a complex system allows for inferring the dynamics which emerge from such individual characteristics at the level of an economic system, such as organizations or the society. Understanding these dynamics is of ultimate importance for the fields of Business and Economics, as it supports organizations and policymakers in anticipating and coordinating behavior and, thereby, helps reduce the amplification of uncertainty potentially resulting from the COVID-19 pandemic.

This position paper almost exclusively focuses on the above-mentioned fields of Computer Science and Information Systems, (Computational) Linguistics, Cognitive Psychology, and Business and Economics. In order to get a full(er) and deep(er) understanding of the dynamics emerging from infodemics, the integration of further scientific fields, such as Sociology, Ethics, Cultural Studies, and the Legal Sciences is strongly needed. Thus, aside from highlighting the importance of the fields currently in focus for research on infodemics, we conclude with a call to action for strongly interdisciplinary research on this topic involving related fields as well. Such exceptional times require us to look beyond the horizon of our own discipline and to join forces in order to unfold collective creativity.